\documentclass{article}
\usepackage{waspaa17,amsmath,graphicx,url,times}
\usepackage{microtype}
\usepackage{cite}
\usepackage{tikz}
\usepackage{mathtools}
\usepackage{amssymb}
\usepackage[hidelinks]{hyperref}

\graphicspath{{./figures/}}
\definecolor{annotred}{rgb}{0.59,0.157,0.078}
\definecolor{labelgray}{rgb}{0.2,0.2,0.2}

\def\R{{\mathbb R}}

\def\dict{{P}}
\def\act{{A}}

\def\spec{{V}}

\newcommand{\apspecvar}[1]{{\dict{#1}}}
\def\apspec{{\apspecvar{\act}}}

\def\distsmall{{d}}
\def\objf{{f}}

\def\charfunc{{\chi}}

\newcommand{\ind}[1]{{_{#1}}}

\DeclarePairedDelimiter{\normcore}{\lVert}{\rVert}
\newcommand{\norm}[1]{\normcore[\big]{#1}}
\def\markov{{\mathcal M}}
\def\threshset{{\mathcal T}}

\def\markov{{\mathcal M}}
\def\minact{{a_\text{min}}}
\def\tdvop{{\Delta_D}}

\title{An Augmented Lagrangian Method for Piano Transcription using \newline Equal Loudness Thresholding and LSTM-based Decoding}

\name{Sebastian Ewert,\thanks{This work was funded by EPSRC grant EP/L019981/1.} Mark B. Sandler}
\address{Machine Listening Lab (MLLAB) and Centre for Digital Music (C4DM)\\School of Electronic Engineering and Computer Science\\Queen Mary University of London\\United Kingdom}

\begin{document}

\ninept
\maketitle

\sloppy
\renewcommand{\baselinestretch}{0.935}
\selectfont

\begin{abstract}
A central goal in automatic music transcription is to detect individual note events in music recordings.
An important variant is instrument-dependent music transcription where methods can use calibration data for the instruments in use.
However, despite the additional information, results rarely exceed an f-measure of 80\%.
As a potential explanation,
the transcription problem can be shown to be badly conditioned and thus relies on appropriate regularization.
A recently proposed method employs a mixture of simple, convex regularizers (to stabilize the parameter estimation process) and more complex terms (to encourage more meaningful structure). 
In this paper, we present two extensions to this method.
First, we integrate a computational loudness model to better differentiate real from spurious note detections.
Second, we employ (Bidirectional) Long Short Term Memory networks to re-weight the likelihood of detected note constellations.
Despite their simplicity, our two extensions lead to a drop of about 35\% in note error rate compared to the state-of-the-art.
\end{abstract}

\begin{keywords}
Proximal Methods, Alternating Directions Method of Multipliers, Structured Sparse Coding, Instrument-dependent Transcription.
\end{keywords}

\section{Introduction}
\label{sec:intro}

Automatic music transcription (AMT) is often considered to be a key technology in music processing as it provides a link between the acoustic domain (in the form of audio recordings) and the symbolic music domain (capturing note events and higher level musical concepts) \cite{BenetosDGKK12_AMTGlassCeiling_ISMIR}.
A central component in an AMT system is the detection of individual note events in an audio recording of a piece of music.
However, despite ongoing research since the 1970s \cite{Moorer77_Transcription_CMJ}, the AMT problem remains unsolved in its most general form \cite{BenetosDGKK12_AMTGlassCeiling_ISMIR}, i.e. for an unknown number of instruments of unknown type playing jointly under unknown acoustic conditions. 
In particular, a major challenge is that in music note events are usually highly correlated both in time and frequency -- from a modelling point of view this often results in highly ill-conditioned systems of (non-)linear equations \cite{EwertS16_PianoTranscriptionADMM_TASLP}.

Several families of AMT methods have been proposed, each building on specific strategies to approach the AMT problem, see \cite{KlapuriD06_SPforMusic_BOOK,BenetosDGKK12_AMTGlassCeiling_ISMIR,EwertS16_PianoTranscriptionADMM_TASLP} for an overview.
Currently, most state of the art methods either employ neural networks \cite{KelzDKBA2016_SimpleFramewisePianoTrans_ISMIR,SigtiaBD2016_NNPianoTrans_TASLP} (typically using discriminative modelling) or factorization methods \cite{BenetosEW14_PitchUnpitchedTranscription_ICASSP} (i.e. inference methods in generative models).
Using the piano transcription task as an example, current methods typically yield f-measures (for correctly detected notes) of around 50-70\%, leaving considerable room for improvement. 

A typical approach to increase the transcription accuracy is to include recordings of the instrument to be transcribed in the training material -- this is valid in a variety of scenarios where a calibration phase is possible (e.g. studio or home recordings). We will refer to this problem scenario as \emph{instrument-dependent music transcription}.
However, despite the availability of additional information, the f-measure for many methods improves only slightly to 60-80\% -- this range holds for both discriminative methods \cite{KelzDKBA2016_SimpleFramewisePianoTrans_ISMIR,SigtiaBD2016_NNPianoTrans_TASLP} and generative models \cite{OHanlonNKP2016_NonNegativeGroupSparseTranscription_TASLP,CogliatiDW2017_PianoTransConvolutional_IEEESPL}.
For example, in \cite{KelzDKBA2016_SimpleFramewisePianoTrans_ISMIR} Kelz et al.\ describe the current state-of-the-art based on neural networks (instrument-dependent training) -- the final proposed method achieves an f-measure of $\approx$$80$\%, which is achieved employing an extensive hyper parameter tuning process.

A first idea to improve ill-conditioned problems is to lower the 'noise', which means to keep the patterns used for identification as close as possible to the observations.
In particular, for instruments such as the piano, a note is not a stationary sound but rather evolves in typical formations over time.
Most factorization based methods, however, do not take this temporal progression into account and rather employ pure spectral templates.
The idea in \cite{EwertPS15_DPNMD_ICASSP} is thus to model this note progression employing a graphical model that controls the temporal position in 88 spectro-temporal patterns, each associated with one piano key.
The model is conceptually similar to Non-negative Matrix Deconvolution (NMD) \cite{Smaragdis04_NMD} but employs, in contrast to NMD, patterns of variable length.
There is also a close connection to non-negative factorial HMMs (NFHMM) \cite{MysoreS12_VariationalNFHMM_ICML,BenetosCW13_EfficientSIModel_IWMLM} %
-- the main difference in \cite{EwertPS15_DPNMD_ICASSP} being in the use of a specialized parameter estimation process to enable the use of 88 parallel Markov processes.

Overall, the system presented in \cite{EwertPS15_DPNMD_ICASSP} models the piano sound production process quite closely. Yet, this was not reflected in the evaluation results, with f-measure values around $80$\% on a standard dataset (MAPS \cite{EmiyaBD10_MultipitchEstimation_TASLP}).
A detailed analysis conducted in the context of \cite{EwertS16_PianoTranscriptionADMM_TASLP} revealed that the signal model can be used to yield a  higher transcription accuracy. However, for numerical reasons, the underlying parameter estimation process used in \cite{EwertPS15_DPNMD_ICASSP} was biased towards
specific local minima of an objective function that are likely to cause misdetections.
The design goal in \cite{EwertS16_PianoTranscriptionADMM_TASLP} was thus to use a signal model similar to \cite{EwertPS15_DPNMD_ICASSP} but to replace the entire parameter estimation process. The resulting method consecutively switches from simple, convex regularizers (that stabilize the initial parameter estimation process) to more complex terms (to encourage a more meaningful structure as expressed by a graphical model). As a result, focusing only on the numerical properties of the parameter estimation process, the methods yields f-measure values of $95$\%.

While this is a step forward, the performance is still not high enough for all relevant applications -- intuitively, still 5 in 100 notes are not correctly detected. The contribution in this paper is to investigate additional strategies for further increasing this accuracy. Such a high accuracy is particularly important in education systems that are used to give students feedback on their mistakes \cite{BenetosKD12_ScoreInformedAMT_EUSIPCO,EwertWMS16_ScoreDeviation_ISMIR}. In this context, a manual analysis revealed two important sources of errors: notes played with a low intensity and additive noise (moving chairs, coughing).
With respect to the first problem, the system in \cite{EwertS16_PianoTranscriptionADMM_TASLP} employs a single threshold for all pitches to differentiate real from spurious notes. Using only a single value, however, is problematic as loudness perception depends on the frequency. Therefore, a first simple extension is to make this threshold pitch-dependent. Here, as we will see, using simple schemes such as equal loudness contours that are based on sinusoidal tones did not give an advantage -- instead we incorporated a method based on the Glasberg-Moore model for complex, non-stationary sounds \cite{GlasbergM2002_ModelLoudnessTimeVarying_JAES}.

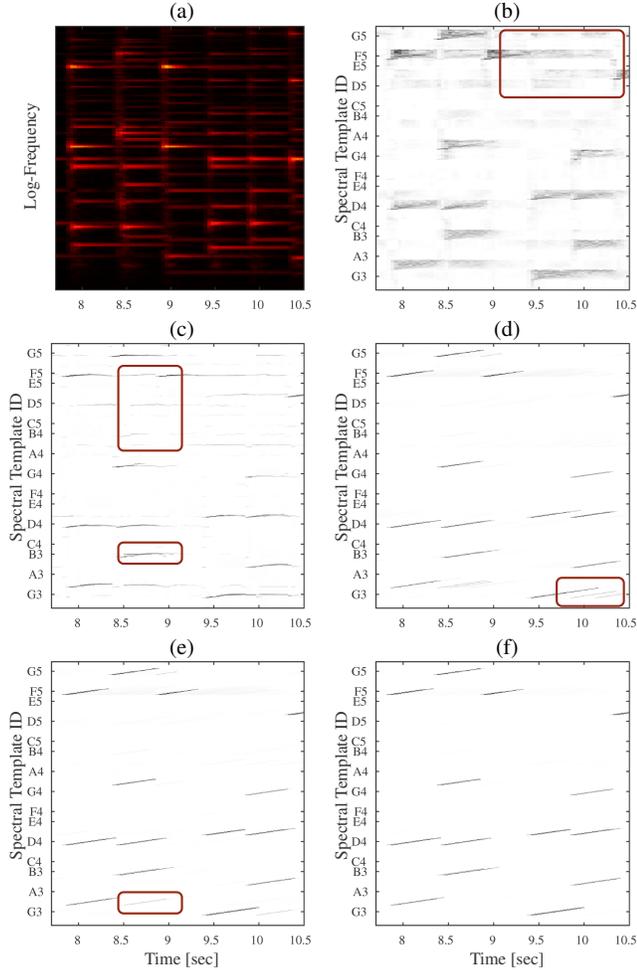
\begin{figure}[t]
    \begin{tikzpicture}[scale=0.94, every node/.style={transform shape}]
			\begin{scope}[shift={(4.6,9)}]
      		\node[anchor=south west] (label) at (0.4,0){\includegraphics[width=3.6cm,height=4cm]{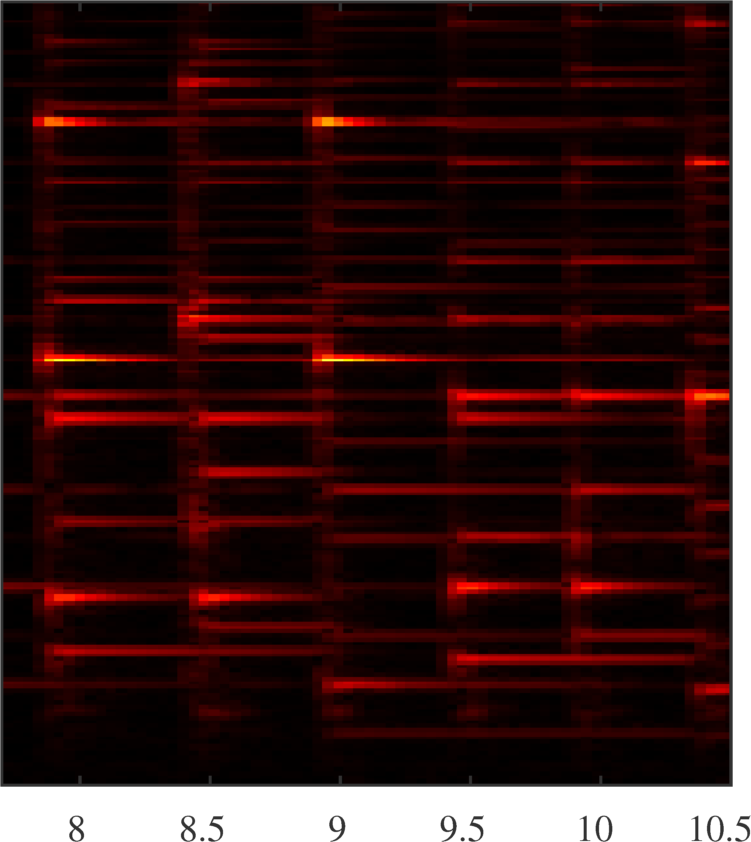}};
			\node[anchor=south west,rotate=90] at (0.4,1.4) {\scriptsize \textcolor{labelgray}{Log-Frequency}};
			\node (label) at (2.3,4.3) {\small (a)}; 
			\end{scope}
			\begin{scope}[shift={(9.2,9)}]
      		\node[anchor=south west] (label) at (0,0){\includegraphics[width=4cm,height=4cm]{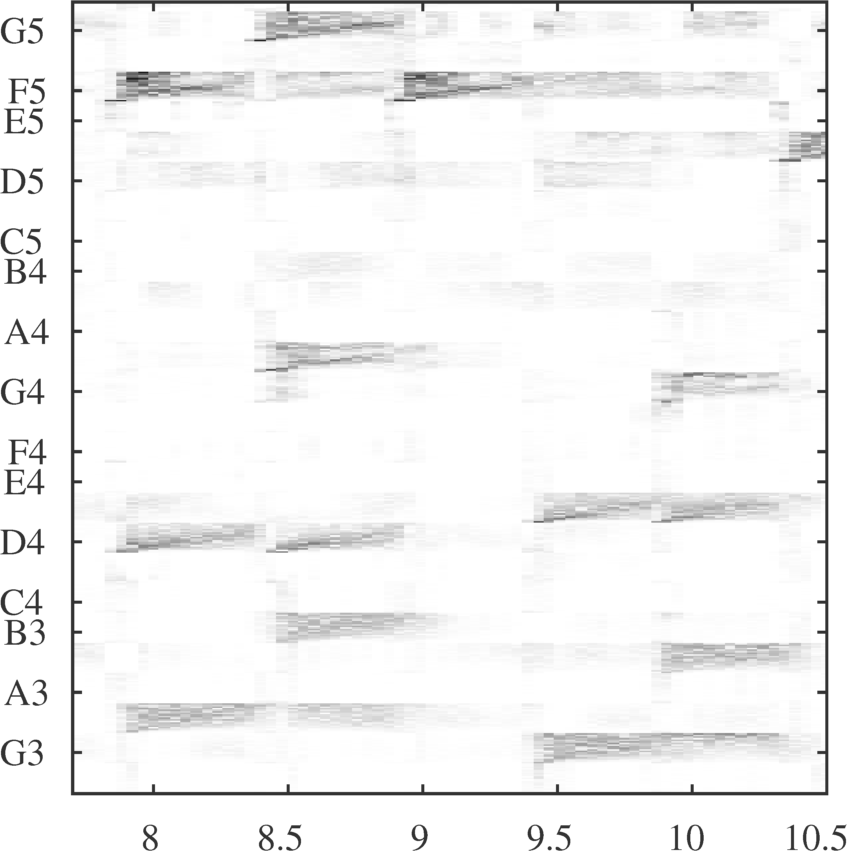}};
			\node[anchor=south west,rotate=90] at (0.2,1.2) {\scriptsize \textcolor{labelgray}{Spectral Template ID}};
			\node (label) at (2.3,4.3) {\small (b)}; 
			\draw[annotred,thick,rounded corners=2pt] (2.2,3.1) rectangle (3.95,4.05);
			\end{scope}
			\begin{scope}[shift={(4.6,4.5)}]
      		\node[anchor=south west] (label) at (0,0){\includegraphics[width=4cm,height=4cm]{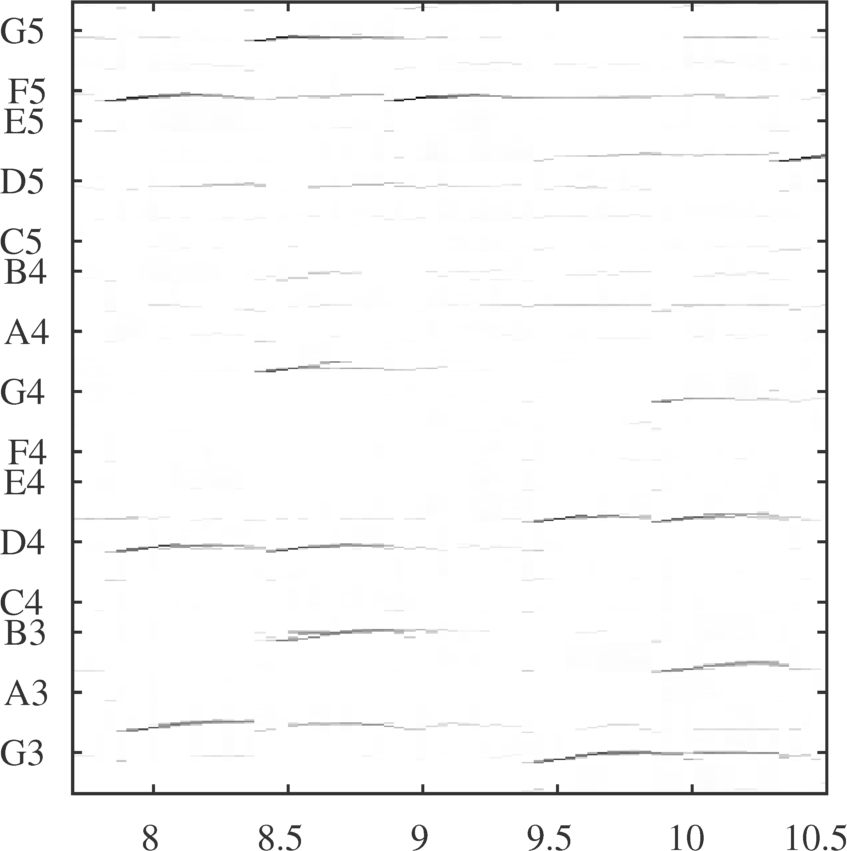}};
			\node[anchor=south west,rotate=90] at (0.2,1.2) {\scriptsize \textcolor{labelgray}{Spectral Template ID}};
			\node (label) at (2.3,4.3) {\small (c)}; 
			\draw[annotred,thick,rounded corners=2pt] (1.4,1) rectangle (2.3,1.3);
			\draw[annotred,thick,rounded corners=2pt] (1.4,2.6) rectangle (2.3,3.8);
			\end{scope}
			\begin{scope}[shift={(9.2,4.5)}]
      		\node[anchor=south west] (label) at (0,0){\includegraphics[width=4cm,height=4cm]{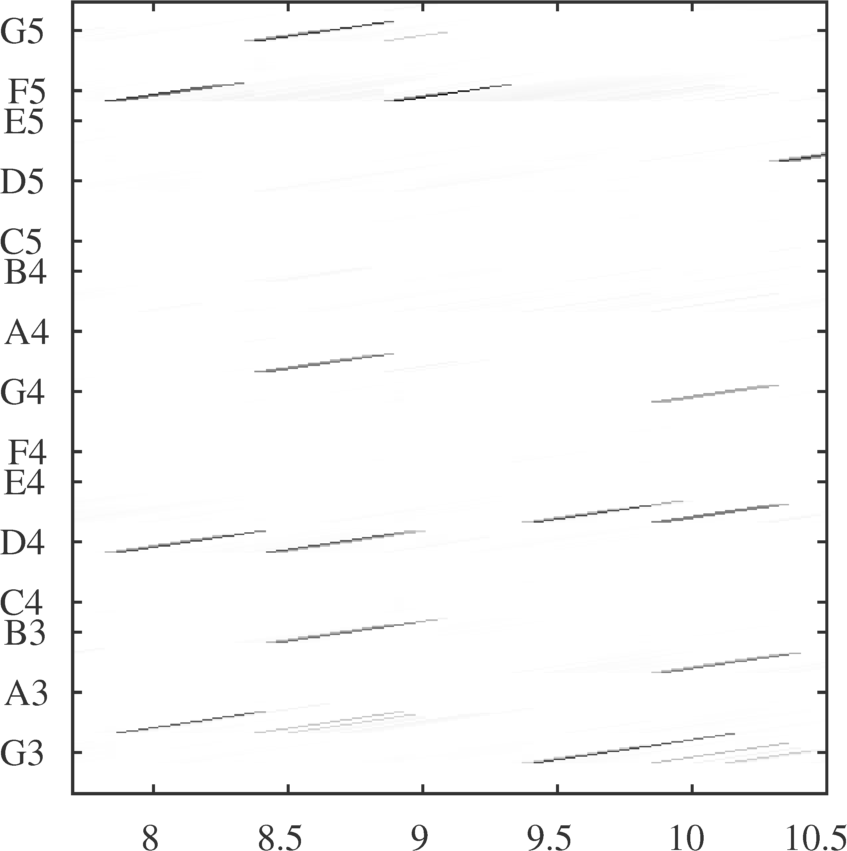}};
			\node[anchor=south west,rotate=90] at (0.2,1.2) {\scriptsize \textcolor{labelgray}{Spectral Template ID}};
			\node (label) at (2.3,4.3) {\small (d)}; 
		  	\draw[annotred,thick,rounded corners=2pt] (3.0,0.4) rectangle (3.95,0.8);
			\end{scope}
			\begin{scope}[shift={(4.6,0)}]
      		\node[anchor=south west] (label) at (0,0){\includegraphics[width=4cm,height=4cm]{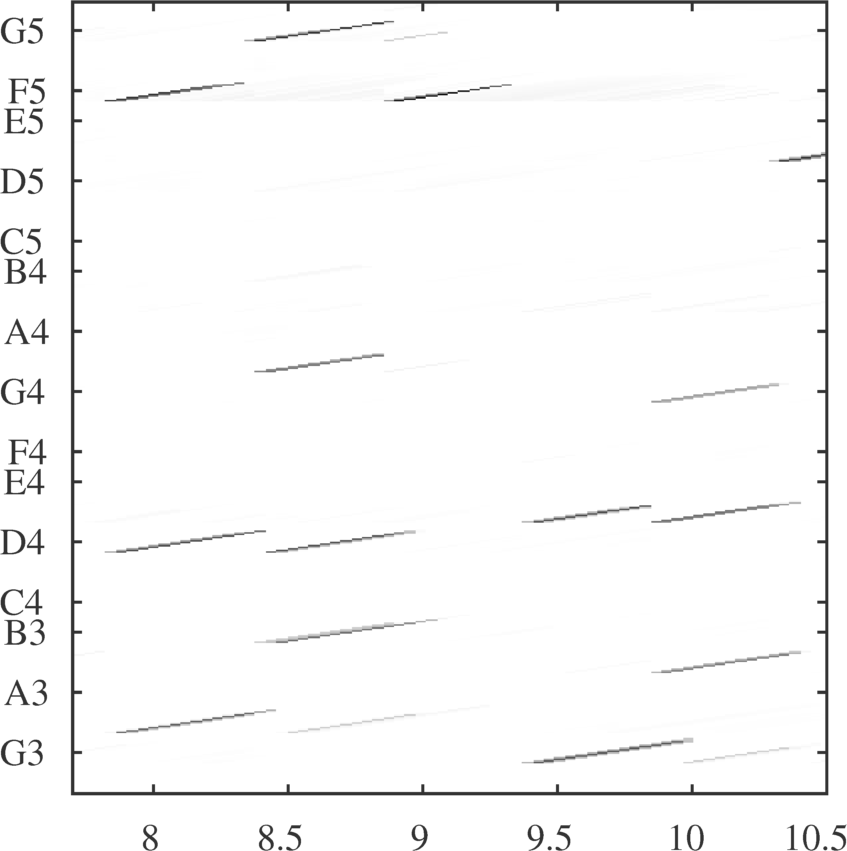}};
			\node[anchor=south west,rotate=90] at (0.2,1.2) {\scriptsize \textcolor{labelgray}{Spectral Template ID}};
			\node at (2.3,-0.1) {\scriptsize \textcolor{labelgray}{Time [sec]}};
			\node (label) at (2.3,4.3) {\small (e)}; 
			\draw[annotred,thick,rounded corners=2pt] (1.4,0.55) rectangle (2.3,0.85);
			\end{scope}
			\begin{scope}[shift={(9.2,0)}]
      		\node[anchor=south west] (label) at (0,0){\includegraphics[width=4cm,height=4cm]{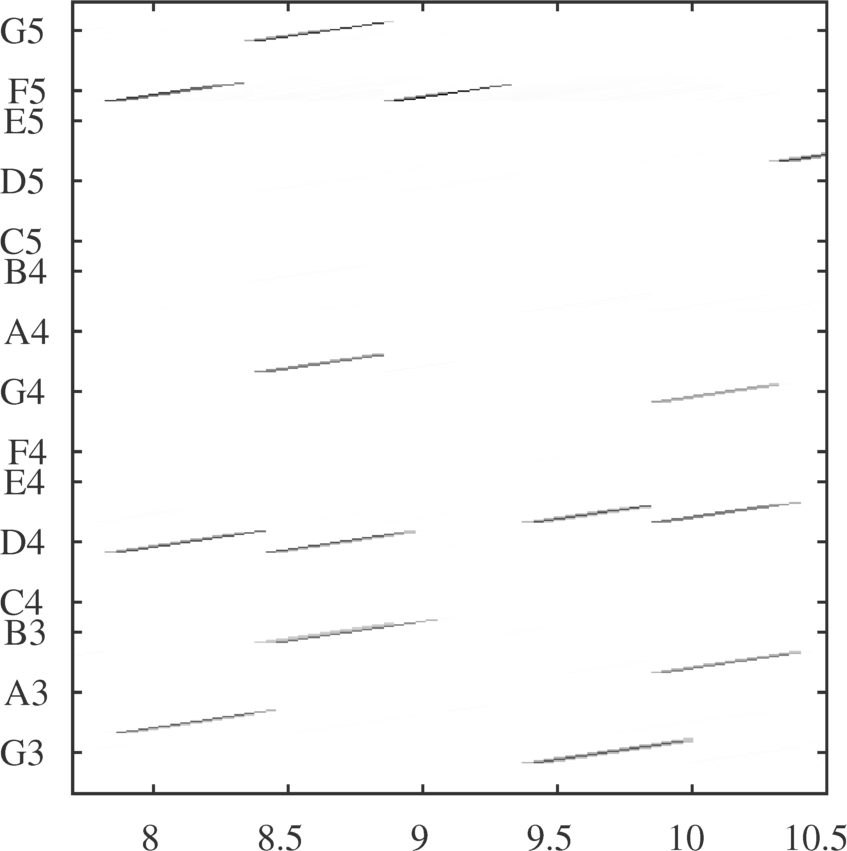}};
			\node[anchor=south west,rotate=90] at (0.2,1.2) {\scriptsize \textcolor{labelgray}{Spectral Template ID}};
			\node at (2.3,-0.1) {\scriptsize \textcolor{labelgray}{Time [sec]}};
			\node (label) at (2.3,4.3) {\small (f)}; 
			\end{scope}
    \end{tikzpicture}
\vspace{-6mm}
\caption{Illustration of the effect of using different combinations of regularizers. \textbf{(a)} Log-frequency spectrogram of a recording of Chopin's Nocturne No.2 (Op.~9). \textbf{(b)-(g)} Activity tensor estimated using different combinations of regularizers, see text for details.}
\label{fig:IlluRegul}
\end{figure}

To deal with additive noise, we can exploit that additive sounds as described above typically lead to activations that are harmonically unrelated to the music. That means we need a measure for how likely a certain constellation of notes is. This could be implemented using an HMM -- however, since we are modeling constellations of notes, the corresponding state-space would at least have a size of $2^{88}$ (one state for each combination of active notes), which is practically infeasible. However, such large, complex joint distributions have recently been successfully approximated using neural networks \cite{BoulangerBV12_DeepRNN_ICML}\cite{SigtiaBD2016_NNPianoTrans_TASLP}. Therefore, as a second extension, we investigate here combining the method proposed in \cite{EwertS16_PianoTranscriptionADMM_TASLP}, which is adaptable to new acoustical conditions with minimal effort, with long short term memory (LSTM) neural networks for decoding, which essentially provide a simple musical language model on top. 

The remainder is organized as follows.
In Section~\ref{sec:proposedMethod}, we describe our proposed extensions in more detail and report in Section~\ref{sec:experiments} on our evaluation results. We conclude in Section~\ref{sec:conclusions} with a prospect on future work.

\section{Proposed Method}
\label{sec:proposedMethod}

\subsection{Signal Model}

Before we discuss our proposed extensions, we begin with a short summary of the model as presented in \cite{EwertS16_PianoTranscriptionADMM_TASLP} and refer there for many of the details.
The core of our signal model corresponds to a tensor product modeling a time-frequency representation $V \in \R_{\ge 0}^{M,N}$ of a recording to be transcribed:
\begin{equation}
\spec\ind{m,n} \approx (\apspec)\ind{m,n} := \sum_k \sum_\ell \dict\ind{m,\ell,k} \cdot \act\ind{k,\ell,n}.
\label{eq:mainModel1}
\end{equation}
The \emph{pattern dictionary tensor} $\dict \in \R_{\ge 0}^{M \times L \times K}$ contains $K$ spectro-temporal patterns, each consisting of $L$ frames. Here, $K=88$ corresponds to the number of keys on a piano. That means each column $\dict\ind{:,\ell,k} \in \R_{\ge 0}^{M}$ for fixed $\ell$ and $k$ contains a single \emph{spectral template}; here we used the slicing notation \emph{:} to refer to all elements in an index dimension. Each pattern $\dict\ind{:,:,k}$ corresponds to a recording of a single note. 
The \emph{activity tensor} $\act \in \R_{\ge 0}^{K \times L \times N}$ encodes the activity of each template in each frame.

This basic signal model is relatively free and thus requires strong regularization. The following objective function employs several of the regularizers as proposed in \cite{EwertS16_PianoTranscriptionADMM_TASLP}:
\begin{align}
\objf(\act) := & \quad \sum_{m,n}\distsmall(\spec\ind{m,n},(\apspec)\ind{m,n}) \label{eq:KL}\\
& + \charfunc_{\R_{\ge 0}^{K\times L \times N}}(\act) \label{eq:NN}\\
& + \lambda_1 \norm{\act}_1 \label{eq:L1}\\
& + \lambda_2 \norm{\tdvop[\act]}_1 \label{eq:TDV}\\
& + \charfunc_\markov(\act) \label{eq:Markov}\\
& + \charfunc_\threshset(\act) \label{eq:Thresh}
\end{align}
We illustrate the idea behind each term in Fig.~\ref{fig:IlluRegul}. In Fig.~\ref{fig:IlluRegul}a we see a time-frequency representation of the recording to be transcribed. Fig.~\ref{fig:IlluRegul}b-f show activity tensors $\act$ obtained using different subsets of the terms (\ref{eq:KL})--(\ref{eq:Thresh}). For illustrative purposes, each $\act$ is flattened out by placing the slices $\act\ind{k,:,:}$ vertically on top of each other.
With $\distsmall(a,b) := a \cdot \log\left(\frac{a}{b}\right) - a + b$ for $a,b> 0$ term~(\ref{eq:KL}) is a data fidelity term using the generalized Kullback-Leibler divergence. Term~(\ref{eq:NN}) encourages non-negativity of $\act$, where $\charfunc_S$ is the characteristic function for some set $S$ with $\charfunc_S(x)=0$ if $x \in S$ and $\charfunc_S(x)=\infty$ otherwise. Fig.~\ref{fig:IlluRegul}b shows the result of using only terms (\ref{eq:KL})--(\ref{eq:NN}): As discussed in \cite{EwertS16_PianoTranscriptionADMM_TASLP}, the corresponding $\act$ is blurred and noisy -- due to the structure of $\dict$ we would expect diagonal lines that start at the position of note onsets. A transcription based on such a representation is likely to contain a larger number of errors.

To obtain more meaningful activations, term~(\ref{eq:L1}) introduces a sparsity inducing $\ell_1$ regularizer. As shown in Fig.~\ref{fig:IlluRegul}c, the results indeed clear up. However, instead of diagonal lines, we rather see horizontal lines. This is caused by using single note patterns in $\dict$ whose individual templates are not normalized, i.e. we preserve the characteristic energy decay in the pattern. This causes here, however, that only the energy-rich, first templates in a pattern are activated. This activation of 'wrong' templates causes residual energy and thus spurious activity. As a remedy, term~(\ref{eq:TDV}) discourages changes along diagonals using an anisotropic variant of the total variation operator. Here, $\tdvop$ is essentially a simple high-pass filter along the diagonals as introduced in \cite{EwertS16_PianoTranscriptionADMM_TASLP}. As shown in \cite{EwertS16_PianoTranscriptionADMM_TASLP}, terms~(\ref{eq:KL})--(\ref{eq:TDV}) are jointly convex in $\act$ -- when used to obtain a first initialization for $\act$ the convexity improves the robustness of the method considerably.

However, while convexity improves numerical stability, it often limits the expressiveness of terms. To improve upon remaining problems, additional non-convex terms are added. One remaining problem can be see in the G3 activations around 10 seconds (Fig.~\ref{fig:IlluRegul}d): The G3 is activated twice, which is physically impossible and can lead to estimation errors. As a countermeasure, term~(\ref{eq:Markov}) uses the characteristic function with a very specific set $\markov$. This set contains only tensors $\act$ whose activations encode states in a specific graphical model. The model essentially encodes that a note has a minimum length and how it can progress in time. Including term~(\ref{eq:Markov}) resolves the concurrency issues (Fig.~\ref{fig:IlluRegul}e). Due to space restrictions, we refer to \cite{EwertS16_PianoTranscriptionADMM_TASLP} for details.

A final problem is visible in Fig.~\ref{fig:IlluRegul}e: a weak, incorrect activation of G\#3 around 8.5 seconds (octave error). If the note energy is distributed across several weak activations, the correct activation can fall below the detection threshold. For this reason, term~(\ref{eq:Thresh}) specifies with $\threshset := \big( [\minact,\infty) \cup \{0\} \big)^{K \times L \times N}$ that activations have to be zero or greater than $\minact$. This way, low intensity energy is 'pulled' into the main activation which in extreme cases pushes the activation above the detection boundary. In this context, see also \cite{BlumensathD2010_IterativeHardThreshL0_JSTSP} for a connection between hard thresholding and hard $\ell_0$ sparsity.

\subsection{Parameter Estimation using the Augmented Lagrangian}

To obtain a meaningful $\act$, we need to find a minimizing argument to our objective function $\objf$.
However, such a function is difficult or even impossible to minimize with classical gradient or Newton-type optimization methods.
It contains highly non differentiable terms, terms that yield infinity as value and strongly non-convex terms. 
In this context, Augmented Lagrangian methods have been found to be of high interest. 
In particular, the variant \emph{Alternating Directions Method of Multipliers (ADMM)} \cite{BoydPCPE11_ADMM_Book} 
provides a scheme to split up the
objective function, minimizing the terms individually and still
provides convergence guarantees for the entire objective. As a
result it is not only useful for complex objective functions as
in our case but also in big data scenarios, as ADMM's splitting
and merging operations fit perfectly into distributed computing
schemes like Map-Reduce. Due to space constraints we refer to \cite{EwertS16_PianoTranscriptionADMM_TASLP} for more details on ADMM and minimizing the objective function $\objf$. 

\subsection{Thresholding based on Glasberg-Moore Model}

Before we describe a first extension to the method introduced in \cite{EwertS16_PianoTranscriptionADMM_TASLP}, we identify a potential weakness.
In particular, the hard thresholding introduced by term~(\ref{eq:Thresh}) was found to be particularly useful to improve the detection of low intensity notes, i.e. those close to the decision boundary. In \cite{EwertS16_PianoTranscriptionADMM_TASLP}, the corresponding threshold $\minact$ was derived from user input as this threshold depends on the recording level. More precisely, the user is asked to provide an example of a note having the lowest intensity to be expected in a recording session (during the evaluation this was one value for the entire dataset and not specific to recordings). The system processes this low-intensity recording and sets $\minact$ using the computed activation values.

\begin{figure}[t]
\vspace{-1mm}
\centering
\centerline{
\begin{tikzpicture}[scale=0.94, every node/.style={transform shape}]
      		\node[anchor=south west] (label) at (0,0){\includegraphics[width=6cm]{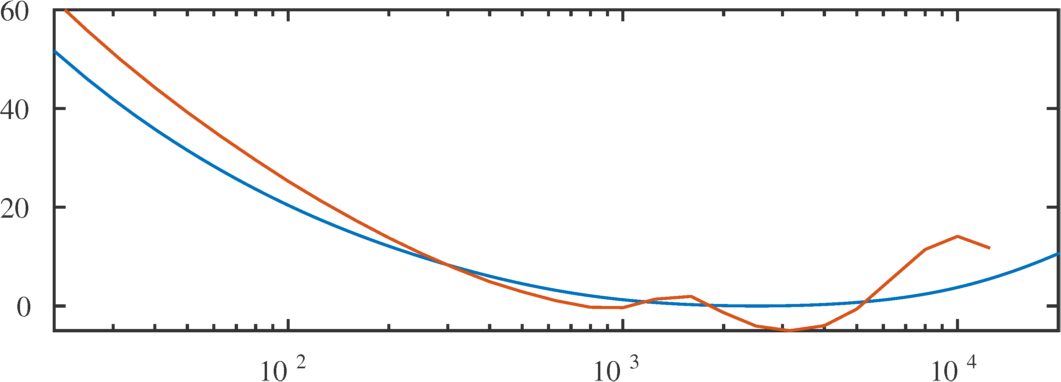}};
			\node[anchor=south west,rotate=90] at (0.1,0.9) {\scriptsize \textcolor{labelgray}{Gain dB}};
			\node at (3.5,-0.1) {\scriptsize \textcolor{labelgray}{Frequency [Hz]}};
\end{tikzpicture}}
\vspace{-4mm}
\caption{Equal loudness curves: Inverted A-weighting (Blue) and Fletcher et al. for 35 Phon (Orange).}
\label{fig:eqloud}
\end{figure}

To keep the user effort minimal, the system employs only a single low-intensity note. While including this type of thresholding in the optimization procedure led to measurable improvements, there is a problem: the perception of loudness is frequency dependent.
That means, if the low-intensity note has a high pitch, its energy is likely to be different from a note having the same perceived loudness but with lower pitch. In other words, an energy-based threshold chosen based on one pitch is likely to be incorrect for another. 
Therefore, in a first extension we make the threshold $\minact$ pitch dependent, without increasing the user effort.

A first idea to implement this change is to set the threshold based on equal loudness curves: Fig.~\ref{fig:eqloud} shows the widely used inverted A-weighting and Fletcher curves \cite{FastlZ07_psychoacoustics_BOOK}.
More precisely, we start by normalizing all single note recordings (that are used to create the pattern tensor $\dict$) to have the same root-mean-square (RMS) energy. Then, for each note, we calculate the difference in dB between the equal loudness value for that note and the one for the low-intensity note -- for this lookup, we use the fundamental frequency associated with each note. Using this difference, we can  derive an individual threshold for each pitch (which then hopefully corresponds to the energy level for a note of that pitch having the same loudness as the low-intensity note).
Unfortunately, this procedure led to virtually no improvement in the results (f-measure improved by 0.2). There was no difference between the A-weighting and the Fletcher curve, which is not surprising given their similarity, compare Fig.~\ref{fig:eqloud}. 

A possible reason could be that there simply are not many low-intensity notes in our evaluation dataset and thus such a measure cannot have a stronger effect. Alternatively, the new thresholds might simply not be meaningful enough as both curves were based on listening tests involving stationary, sinusoidal sounds, while piano notes are harmonic and non-stationary.
To test this hypothesis we employed a more complex loudness model as proposed by Glasberg and Moore \cite{GlasbergM2002_ModelLoudnessTimeVarying_JAES}, which was designed to provide a better fit to complex, non-stationary sounds.
To this end, we derived for each RMS-normalized note a scaling factor such that the scaled note has the same perceived loudness as a reference note (C4 in our case) -- loudness was measured as the local maximum over the entire note duration (Note: All normalized note recordings were pre-scaled to an assumed playback level of 30 db-SPL for the measurement).
We can then use these scalers to convert the threshold obtained from the provided low-intensity note recording to all remaining piano keys.
We will report on the results for this variant in Section~\ref{sec:experiments}.

\subsection{LSTM-based Decoding}

A second occasional problem we observed stems from non-musical interferences, including mechanical sounds from the instrument or breathing sounds.
These sometimes led to spurious activations, in particular, for pitched interferences. 
Most of these activations, however, are not strongly correlated with the music and typically occur as short, out-of-key activations.
Graphical models such as an HMM could be used in this context to smooth over such unusual, musically often irrelevant activations (similar to a language model in speech recognition).
However, even a simple frame-wise model would need to span a space consisting of $2^{88}$ states to model each combination of notes.
While there exist sophisticated pruning techniques for such cases, they tend to be quite complex and often involve considerable trade-offs with respect to approximation quality and runtime performance.
To approximate such complex joint probabilities, there has recently been considerable success using neural networks \cite{BoulangerBV12_DeepRNN_ICML}\cite{EckS2002_BluesWithLSTMNets_WNNSP}\cite{SigtiaBD2016_NNPianoTrans_TASLP} (in the context of symbolic music representations).
Following similar ideas, we have trained long short term memory (LSTM) based recurrent networks to decode the position of onsets in each key, given the activations obtained as above.
While this could have been done with various other architectures as well (e.g. convolutional networks with time context) we chose LSTM networks as our input representation is relatively low-dimensional (convnets are often used to get around the difficulty of training RNNs with high dimensional inputs) and LSTM networks have (theoretically) the model capacity to represent very long temporal dependencies \cite{BoulangerBV12_DeepRNN_ICML}. The latter is achieved by LSTM networks as they provide a more stable gradient flow in the backpropagation-through-time algorithm, which is essentially implemented through gated shortcut connections \cite{EckS2002_BluesWithLSTMNets_WNNSP}. 

As a first step, we divided the activation values associated with each pitch by the corresponding pitch-dependent threshold. 
This way, all activations are normalized to a certain range of values and comparable statistics, which helps with the training process. 
In other words, most instrument and recording specific properties are eliminated from the input and the network can focus on musical aspects, which can be learned independently and do not need to be adapted to new acoustic conditions.
The input for frame $n$ consists of $A\ind{:,1,n}\in \R^{88}$, i.e. the activations for the onset part of each note pattern in $\dict$.
The same representation was used in \cite{EwertS16_PianoTranscriptionADMM_TASLP} for the final onset detection.
We used LSTM networks in two different configurations.
To take the entire recording into account the first configuration uses a bidirectional LSTM network \cite{GravesS2005_FramewisePhonemeBLSTM_NN}.
In such a BLSTM network, one LSTM network operates on the original input sequence and the other one on the reverse sequence.
The two networks are then trained jointly. 
For very long sequences, however, training BLSTM networks typically involves splitting the input sequence into chunks, which is necessary to deal with the limitations in computational resources but leads to questions whether the reversed LSTM network is actually necessary.
Therefore, we trained in a second configuration a uni-directional LSTM.
In this configuration, we simply delay the detection of notes by 400ms to allow the network to peak a little into the future. 

\section{Experiments}
\label{sec:experiments}

To evaluate our proposed extensions, we employed the ENSTDkCl subset of the MAPS collection \cite{EmiyaBD10_MultipitchEstimation_TASLP}, which provides audio recordings of a Yamaha Disklavier and corresponding MIDI-based annotations. 
To evaluate a method, we employ precision (P), recall (R), and F-measure (F) as used in the MIREX evaluation campaigns.
A detected note is considered correct if there is a note in the corresponding ground truth having the same MIDI pitch, with an onset position up to $N$ ms apart from the detected note.
As discussed in \cite{EwertS16_PianoTranscriptionADMM_TASLP}, we set $N$=$100$ to account for some temporal jitter in the ground truth annotations.
Every ground truth note can validate up to one detected note.
For the LSTM networks, we used two layers with 100 units each and a final dense layer containing 88 units with sigmoid activations. For the training we employed a subset of MAPS that was generated using a variety of software synthesizers, i.e. there was no overlap with the test dataset regarding the acoustic conditions. Besides a dropout of 0.5, which we apply only to the non-recurrent connections following \cite{ZarembaSV2014_RNNRegularization_ARXIV}, we employ no other unusual strategies \cite{Bengio2012_PracticalRecommendationsTraining_TOTT}\footnote{We employ Glorot weight initialization \cite{GlorotB2010_UnderstandingDifficultyTraining_AISTATS} and label smoothing \cite{Bengio2012_PracticalRecommendationsTraining_TOTT}.
To normalize the input variance \cite{Bengio2012_PracticalRecommendationsTraining_TOTT}, we measure the variance across both input samples and input dimensions to become invariant against pitch dependent velocity biases in the dataset. Loss is an element-wise cross-entropy. Optimizer is Adam using an initial stepsize set to 1/10 of the default \cite{KingmaB2014_Adam_ARXIV}.}. The results for several methods, including our baseline \cite{EwertS16_PianoTranscriptionADMM_TASLP} and our proposed extensions are given in Table~\ref{tab:dataset_MAPS}.

\begin{table}[t]
\centering
\footnotesize
\begin{tabular}{clccc}
\textbf{Method} & \textbf{P} & \textbf{R} & \textbf{F}\\
\hline
O'Hanlon et al. \cite{OHanlonNKP2016_NonNegativeGroupSparseTranscription_TASLP} & 89 & 77 & 82 \\
Cogliati et al. \cite{CogliatiDW2017_PianoTransConvolutional_IEEESPL} & 80 & 84 & 81 \\
Ewert et al. 2015 \cite{EwertPS15_DPNMD_ICASSP} & 76 & 83 & 79 \\ %
Ewert et al. 2016 (Baseline) \cite{EwertS16_PianoTranscriptionADMM_TASLP} & 96 & 93 & 95 \\ %
Extension 1 & 96 & 95 & 96 \\ %
Extension 1+2 (LSTM) & 97 & 96 & 97 \\ %
Extension 1+2 (BLSTM) & 97 & 96 & 97
\end{tabular}
\caption{Precision, Recall and F-Measure in percent for various methods using the MAPS dataset.}
\label{tab:dataset_MAPS}
\end{table}

The first extension led to an improvement of $0.7$ in f-measure compared to the baseline. Given that there are typically not many notes close to the decision boundary, this might be what can be expected from such a simple extension. We were surprised, however, that this value seemed to be consistently higher than just using equal loudness contours. The use of a more complex loudness model made some difference here.
The LSTM-based decoders improved the results by another $1.0$ in f-measure to an overall improvement of $1.7$. Again, a small improvement but given that the MAPS dataset is very clean and does not contain breathing or similar interferences, we would expect even bigger improvements in actual live recordings. We did not measure a difference in performance between the LSTM and the BLSTM networks, which might indicate that the delayed output solution used in the LSTM network is enough in this specific application scenario.
Overall, the extensions increased the f-measure marginally from $95$\% to $97\%$. While this is an incremental rather than a major step forwards, it means that the expected number of wrong notes in a 100 has gone down from five to three. For feedback systems in education, this can make quite a difference. 

\section{Conclusions}
\label{sec:conclusions}

We presented two extensions improving the accuracy of a state-of-the-art music transcription system. The first extension is based on specifying note-detection boundaries using the Glasberg-Moore loudness model for complex non-stationary sounds. The second extension employs an LSTM network to post-process the output of a dictionary-based method using variable-length spectro-temporal patterns. This way, the capacity to quickly adapt to new acoustic conditions (acoustical model) is combined with a decoder that can focus on music specific aspects such as the likelihood of specific note constellations.
The f-measure increased from $95$\% to $97\%$, corresponding to a drop from five to three incorrect note detections per 100 notes.

\bibliographystyle{IEEEtran}
\bibliography{referencesMusic}

\end{document}